\documentclass[12pt]{article}

\usepackage{amsmath,amsmath,amssymb}
\usepackage{mathrsfs}
\usepackage{multirow}
\usepackage{graphicx}
\usepackage{authblk}
\usepackage{indentfirst}
\usepackage{multicol}  
\usepackage{tabu}
\usepackage{tabularx}
\usepackage{url}
\usepackage{fancyhdr}
\usepackage[numbers]{natbib}
\usepackage{lineno}
\usepackage{makecell}
\usepackage{fancyhdr}
\usepackage{colortbl}
\usepackage{caption}
\usepackage{float}
\usepackage[margin=1in]{geometry} 

\usepackage{setspace} 

\usepackage{hyperref}
\hypersetup{colorlinks=true,linkcolor=blue,anchorcolor=blue,citecolor=blue}

\title{Topological Machine Learning for Protein-Nucleic Acid Binding Affinity Changes Upon Mutation}

\author[1]{Xiang Liu} 
\author[1]{Junjie Wee} 
\author[1,2,3]{Guo-Wei Wei \thanks{Corresponding author: weig@msu.edu}}
\affil[1]{Department of Mathematics, Michigan State University, MI, 48824, USA}
\affil[2]{Department of Electrical and Computer Engineering, Michigan State University, MI 48824, USA}
\affil[3]{Department of Biochemistry and Molecular Biology, Michigan State University, MI 48824, USA}

\date{}

\begin{document}
\maketitle

\paragraph{Abstract} 
Understanding how protein mutations affect protein-nucleic acid binding is critical for unraveling disease mechanisms and advancing therapies. Current experimental approaches are laborious, and computational methods remain limited in accuracy. To address this challenge, we propose a novel topological machine learning model (TopoML) combining persistent Laplacian (from topological data analysis) with multi-perspective features: physicochemical properties, topological structures, and protein Transformer-derived sequence embeddings. 
This integrative framework captures robust representations of protein-nucleic acid binding interactions. To validate the proposed method, we employ two datasets, a protein-DNA dataset with 596 single-point amino acid mutations, and a protein-RNA dataset with 710 single-point amino acid mutations. We show that the proposed TopoML model outperforms state-of-the-art methods in predicting mutation-induced binding affinity changes for protein-DNA and protein-RNA complexes.

  
\newpage
	
\section{Introduction}
Protein-nucleic acid interactions play fundamental roles in a variety of biological activities, including DNA replication, transcription, and translation \cite{ollis1987structural,konig2012protein,zhang2019comprehensive}. Since nucleic acids are highly charged polymers, protein-nucleic acid interactions are largely determined by various intermolecular polar forces, such as hydrogen bonding, dipolar forces, electrostatic interactions, although van der Waals forces and hydrophobic interactions may play a minor role. Even a single amino acid mutation in a protein, such as from a positively charged amino acid residue to a negatively charged one, can disrupt these forces, thereby altering the interaction with nucleic acids \cite{trelsman1989single,luscombe2002protein}. This may result in significant functional impairments, leading to various diseases \cite{kechavarzi2014dissecting,sibanda2017dna,wang2020cancer,agrawal2021frontotemporal}. Therefore, investigating the impact of amino acid mutations on protein-nucleic acid interactions is critical for understanding disease mechanisms and for guiding the development of effective therapeutic interventions.

The strength of protein-nucleic acid interactions is typically quantified by binding affinity $(\Delta G)$ or dissociation constant $(K_d)$. The binding affinity changes from a mutation is commonly computed as $\Delta\Delta G=\Delta G(mutant)-\Delta G(wild$-$type)$. Several experimental methods, including surface plasmon resonance \cite{nguyen2015surface}, isothermal titration calorimetry \cite{velazquez2004isothermal}, and fluorescence resonance energy transfer \cite{hillisch2001recent}, have been developed to measure the binding affinities between proteins and nucleic acids. However, these experimental methods are often time-consuming and labor-intensive, making them unsuitable for large-scale mutational analysis. Consequently, there is a critical need to develop computational approaches for predicting protein-nucleic acid binding affinity changes upon mutation.

Over the past decade, numerous machine learning models have been proposed to predict the protein-nucleic acid binding affinity changes upon mutation. Among these, SAMPDI \cite{peng2018predicting}, SAMPDI-3D \cite{li2021sampdi}, SAMPDI-3Dv2 \cite{rimal2025further}, and PremPDI \cite{zhang2018prempdi} focus on protein-DNA interactions, while PremPRI \cite{zhang2020prempri} and PRA-MutPred \cite{harini2025mutpred} are designed for protein-RNA interactions. Some models, such as mCSM-NA \cite{pires2017mcsm}, PEMPNI \cite{jiang2021systematic}, and PNBACE \cite{xiao2024pnbace}, are for both protein-DNA and protein-RNA interactions. SAMPDI-3Dv2 and PRA-MutPred represent the most recent advancements for protein-DNA and protein-RNA interactions, respectively. The S596 dataset provided by SAMPDI-3Dv2 and the dataset of 710 mutations provided by PRA-MutPred are currently the largest available protein-DNA and protein-RNA datasets that have been used for evaluating models in the prediction of mutation-induced binding affinity changes. Despite these developments, the performance of these models remains modest,  with SAMPDI-3Dv2 and PRA-MutPred achieving PCCs of only 0.65 and 0.70 in the k-fold cross-validation, respectively, calling for novel strategies to enhance prediction accuracy.

Topological data analysis (TDA) is an emerging field that uses topological tools, particularly, persistent homology,  to analyze the shape and structural patterns within data \cite{zomorodian2004computing,epstein2011topological}. In computational biology, TDA was combined with machine learning for protein classification \cite{cang2015topological}. Topological deep learning (TDL) was first introduced by Cang and Wei in 2017 to integrate TDA and deep neural networks \cite{cang2017topologynet}. In the same work, TDL outperformed other competing approaches in protein-ligand binding and protein mutation predictions \cite{cang2017topologynet}. TDL has become the new frontier in rational learning \cite{papamarkou2024position}. 
Most impressively, this approach achieved victories in D3R Grand Challenges, a global competition series in computer-aided drug design \cite{nguyen2019mathematical,nguyen2020mathdl}. Essentially, TDA approaches are extremely effective for intrinsically complex multiscale data, such as those from the biomoleculear structures. 
However, persistent homology has many limitations, including its insensitivity to no topological shape evolution.  Persistent spectral graph, also known as persistent Laplacian, was introduced  to overcome some limitations of persistent homology by Wang et al. \cite{wang2020persistent} Persistent Laplacians not only encode topological information through the harmonic spectra  (i.e., persistent homology), but also capture rich geometric properties via non-harmonic spectra\cite{wei2025persistent}. A fast algorithm was proposed \cite{memoli2022persistent}, and stability analysis \cite{liu2023algebraic} was given to persistent Laplacians, which justifies successful applications, such as protein-ligand binding prediction \cite{meng2021persistent}.  Persistent Laplacians were shown to outperform persistent homology  with over 30 datasets in protein engineering  \cite{qiu2023persistent}. Notably,  persistent Laplacians have enabled the early prediction of emerging dominant SARS-CoV-2 variants, such as Omicron BA.4 and BA.5, two months prior to their announcement by the World Health Organization (WHO) \cite{chen2022persistent}.
 
It worthy to mention that in addition to earlier TDL prediction of protein stability changes upon mutations \cite{cang2017topologynet}, several TDA-based models, such as TopNetTree \cite{wang2020topology}, PerSpect-EL \cite{wee2022persistent}, HCML \cite{liu2022hom}, PTA \cite{liu2023persistent}, and TopNetmAb \cite{chen2022persistent} have demonstrated strong performance in predicting mutation-induced protein-protein binding affinity changes. Despite the success of these approaches in protein-protein interaction studies, TDL or TDA-based methods, particularly those leveraging persistent Laplacians, have not been applied to protein-nucleic acid interactions.

In this work, we build a persistent-Laplacian-based topological machine learning (TopoML) model for predicting mutation-induced protein-nucleic acid binding affinity changes. The TopoML model integrates three distinct types of features: (1) topological features derived from persistent Laplacians, (2) physicochemical features, (3) sequence features by pretrained protein Transformer. These three types of features are concatenated as the input for a gradient boosting tree algorithm to perform the prediction task. We evaluate TopoML on two benchmark datasets: the S596 dataset comprising 596 mutations for protein-DNA interactions, and a dataset of 710 mutations for protein-RNA interactions. The results demonstrate that TopoML outperforms other competing methods in the predictions of both protein-DNA and protein-RNA interactions, highlighting the effectiveness of incorporating topological features for mutation impact prediction.

\section{Results}
\subsection{Overview of the TopoML Model}
\begin{figure}[h]
    \centering
    \includegraphics[width=1\linewidth]{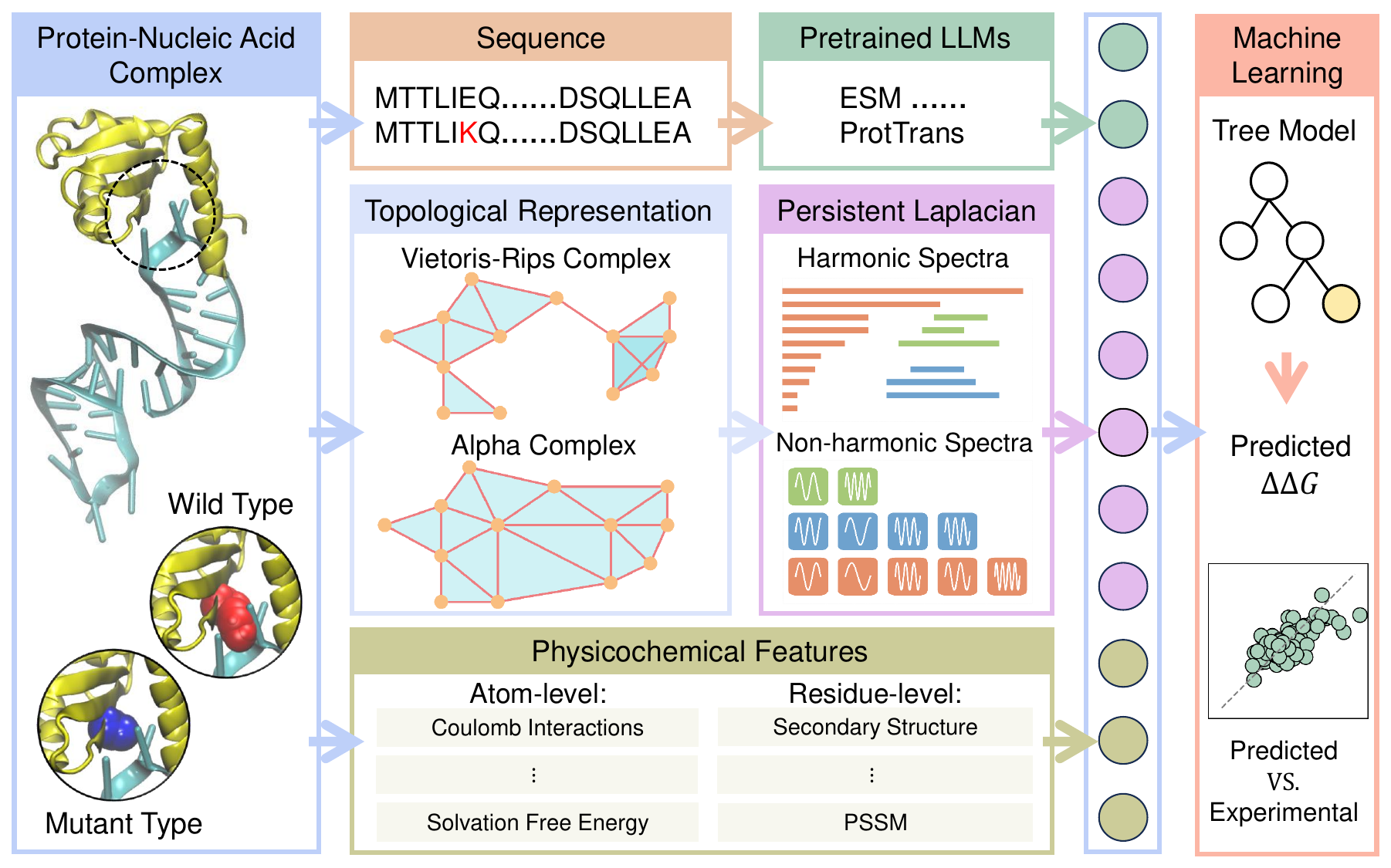}
    \caption{Illustration of TopoML model architecture. For each protein-nucleic acid complex, TopoML extracts a comprehensive set of features and employs a gradient boosting tree algorithm to predict mutation-induced binding affinity changes. The feature generation pipeline comprises three components: sequence features derived from the pretrained protein Transformer models, topological features computed via persistent Laplacian, and the physical and chemical features. These three sets of features are concatenated to form the input of the machine learning algorithm for the prediction task. }
    \label{fig:model}
\end{figure}
The overall workflow of our model is illustrated in Fig. \ref{fig:model}. As a standard machine learning model, TopoML extracts features from protein-nucleic acid complexes and uses a gradient boosting tree algorithm to predict the protein-nucleic acid binding affinity changes upon mutation. The feature generation process consists of three components: sequence features from pretrained protein Transformer, topological features from persistent Laplacians, and physicochemical features. For sequence features, the FASTA sequences of the wild-type and mutant proteins are extracted from the complex and input into the pretrained Transformer models. The derived latent space embeddings are used as the sequence features. 
For topological features, the binding site and mutation site of the protein-nucleic acid before and after mutation are represented as simplicial complexes, and their persistent Laplacians are computed to generate the features. The harmonic spectra of persistent Laplacians are exactly the persistent homology information that characterizes the topological structure information while the non-harmonic spectra encodes more geometric information of the protein-nucleic complex. For physicochemical features, we consider the atom-level properties such as partial charge, electrostatic solvation free energy, and Coulomb interactions, as well as residue-level properties such as mutation site neighbor amino acid composition, p$K_a$ shifts, and additional physicochemical properties. These three types of feature embeddings are concatenated to form the input feature vector for the machine learning algorithm (gradient boosting tree \cite{friedman2001greedy}) to predict the mutation-induced binding affinity changes.

\subsection{Evaluation of the Model}

\subsubsection{Datasets}
Both protein-DNA and protein-RNA complexes are considered to study the effects of the amino acid mutations on binding affinities. For protein-DNA interactions, we use the S596 dataset introduced in a recent work \cite{rimal2025further}. This dataset was systematically curated from the ProNAB \cite{harini2022pronab}, ProNIT \cite{kumar2006protherm}, and dbAMEPNI \cite{liu2018dbamepni} databases, and contains 596 single-point amino acid mutations with experimentally measured $\Delta\Delta G$. For protein-RNA interactions, we consider the dataset from PRA-MutPred \cite{harini2025mutpred}, which was curated from the ProNAB database, and includes 710 single-point amino acid mutations with experimentally determined $\Delta\Delta G$. These two datasets are, to the best of our knowledge, the largest known protein-DNA and protein-RNA datasets that have been used for evaluating models on the prediction of mutation-induced binding affinity changes.

\subsubsection{Evaluation Protocols}
Pearson correlation coefficient (PCC), mean absolute error (MAE), and root mean square error (RMSE) are employed as evaluation metrics to facilitate a fair comparison with benchmark methods reported in the literature \cite{schymkowitz2005foldx,pires2017mcsm,zhang2020prempri,jiang2021systematic,xiao2024pnbace,harini2025mutpred,rimal2025further}. To enhance the reliability of the results, we independently repeated the process 100 times and used the average value as the final performance of our model.

\subsubsection{Performance on Protein-RNA Interactions}

For protein-RNA interactions, the dataset contains 710 single-point amino acid mutations from 134 protein-RNA complexes. Among these mutations, 595 are used as the training set, while the remaining 115 constitute the testing set. Following the strategy employed by PRA-MutPred \cite{harini2025mutpred}, we expand the training set by incorporating reverse mutations, obtained by reversing the mutations and assigning the opposite $\Delta\Delta G$ values as labels. Consequently, we get a training set of 1190 mutations and a test set of 115 mutations.

\begin{figure}[H]
    \centering
    \includegraphics[width=1\textwidth]{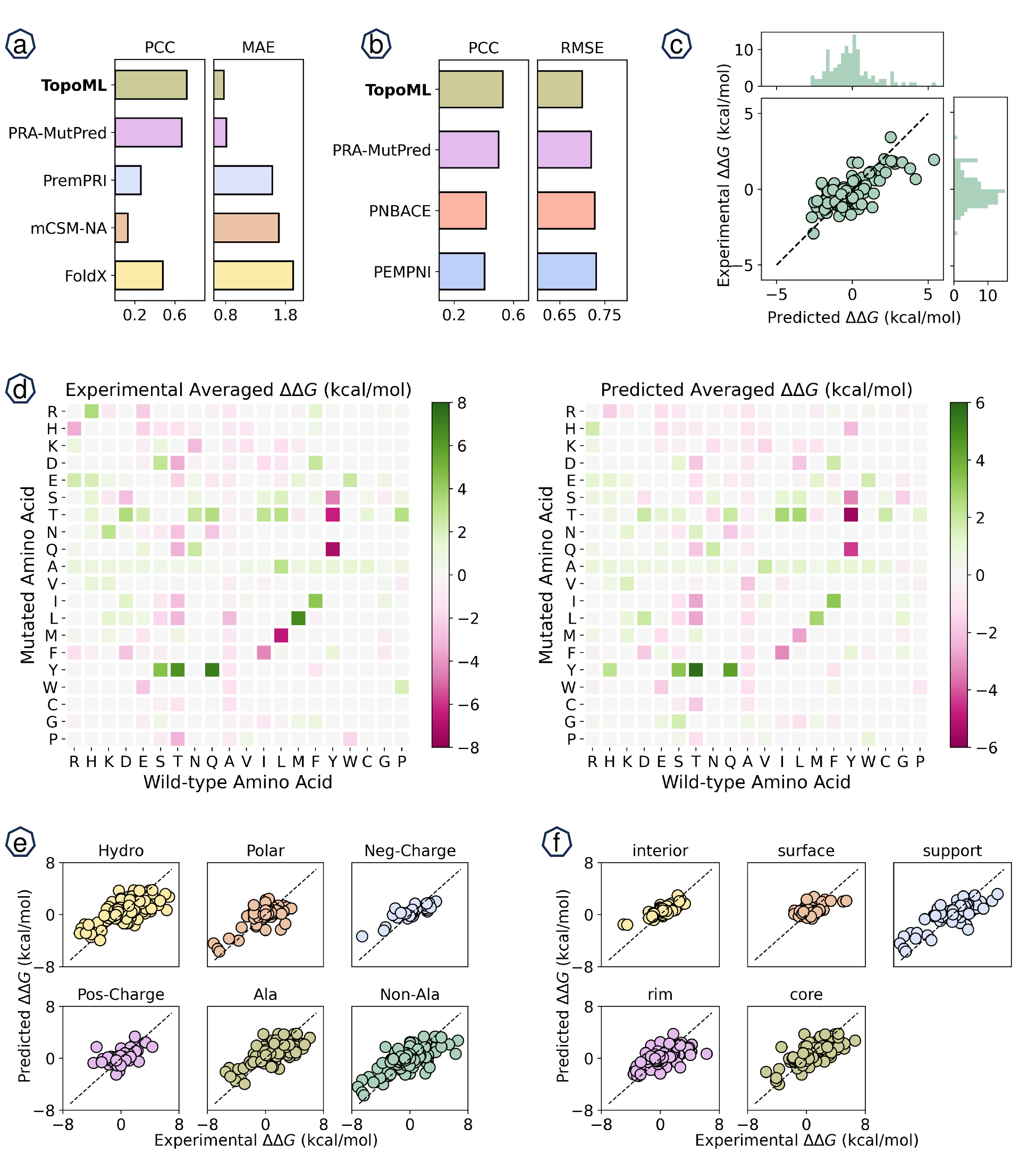}
    \caption{Illustration of model performance for predicting protein-RNA binding free energy changes upon mutation in the S710 dataset. \textbf{(a)}: Performance comparison between our model and existing methods on the test set of 115 mutations. \textbf{(b)}: Performance comparison on the MPR79 test set of 79 mutations, following the protocol used in PRA-MutPred. \textbf{(c)}: Correlation between experimental and predicted $\Delta\Delta G$ values for the test set of 115 mutations. \textbf{(d)}: Residue-residue matrix comparing the averaged experimental and predicted mutation-induced $\Delta\Delta G$ values for the whole dataset of 710 mutations. \textbf{(e)}: Model performance across different mutation types based on residue physicochemical properties. \textbf{(f)}: Model performance across different structural regions defined by relative accessible solvent area (rASA).  }
    \label{fig:protein-rna-result}
\end{figure}

\paragraph{Comparison with Existing Methods}
We evaluate the performance of our model in the test set and compare it with existing methods. The performance comparison is shown in Fig. \ref{fig:protein-rna-result}\textbf{a}. Our model can achieve the best results. Specifically, the FoldX model yields a PCC of 0.48 and an MAE of 1.93 kcal/mol. The mCSM-NA model achieves a PCC of 0.13 with an MAE of 1.69 kcal/mol. The PremPRI model has a PCC of 0.26 and an MAE of 1.58 kcal/mol. PRA-MutPred, the method originally constructed the dataset, reports a PCC of 
\noindent 0.67 and an MAE of 0.81 kcal/mol.  In contrast, our model achieves a PCC of 0.72 and an MAE of 0.77 kcal/mol. A direct comparison between the experimental and predicted $\Delta\Delta G$ values of our model in the test set is shown in Fig. \ref{fig:protein-rna-result}\textbf{c}.

In addition to these models, we also compare our method with PEMPNI, an energy-based model, and its updated version, PNBACE, both of which are designed to predict protein-nucleic acid binding affinity changes upon mutation. These models rely on computationally intensive energy-based features. Following the evaluation strategy of PRA-MutPred, we adopt the test set MPR79, used in the PNBACE study, for a fair comparison, and remove any overlapping mutations from our training set to avoid data leakage. Our model is trained on the remaining data and evaluated on the MPR79 test set. The performance comparison is shown in Fig. \ref{fig:protein-rna-result}\textbf{b}. Our model continues to outperform all baselines. Specifically, PEMPNI has a PCC of 0.41 and an RMSE of 0.73 kcal/mol, while PNBACE reports a PCC of 0.42 with an RMSE of 0.73 kcal/mol. The PRA-MutPred model obtains a PCC of 0.50 and an RMSE of 0.72 kcal/mol. In contrast, our model achieves the best performance with a PCC of 0.53 and an RMSE of 0.70 kcal/mol.

To further evaluate the robustness of our model, we conduct a random 10-fold cross-validation on the full set of 710 mutations. Our model achieves an average PCC of 0.721 and an MAE of 0.785 kcal/mol, outperforming PRA-MutPred, which reports a PCC of 0.70 and an MAE of 0.90 kcal/mol under the same setting.

\paragraph{Performance Analysis Based on Mutation Types}
Using the predictions from the 10-fold cross-validation, we compare the average experimental and our predicted $\Delta\Delta G$ values across different mutation types. As shown in Fig. \ref{fig:protein-rna-result}\textbf{d}, a residue-residue matrix is used where the $x$-axis represents the wild-type residues and the $y$-axis represents the mutated residues. The entry at row of residue `-' and column of residue `+' is the average $\Delta\Delta G$ of all mutations with wild type `+' and mutant type `-'. In the matrix representation, the reverse mutation of each mutation is also considered, the $\Delta\Delta G$ for the reverse mutation, i.e. from mutated type to wild-type, is set to be the opposite value. Consequently, the residue-residue matrix is antisymmetric. It can be seen that the predicted matrix closely mirrors the patterns observed in the experimental data, indicating that our model effectively captures mutation-type-specific trends in binding affinity changes. Although our model shares a similar pattern in average binding affinity changes compared to the experimental data, the variance of its predictions is generally lower, which requires further investigation. One interesting observation is that all mutations to alanine exhibit positive $\Delta\Delta G$ values. A possible explanation is that alanine, with its small and non-polar side chain, is often unable to compensate for the loss of favorable interactions, such as hydrogen bonds and electrostatic contacts that are originally contributed by larger or charged residues. This leads to a reduction in binding affinity and consequently results in positive $\Delta\Delta G$ values. 

Furthermore, we group the mutation residues into different categories: hydrophobic, polar, negatively-charged, positively-charged, alanine, and non-alanine. The comparison between predicted values and experimental ones for each category is shown in Fig. \ref{fig:protein-rna-result}\textbf{e}. Specifically, the PCC (MAE) values are 0.716 (0.728 kcal/mol), 0.73 (0.943 kcal/mol), 0.87 (0.957kcal/mol), 0.469 (1.0kcal/mol), 0.709 (0.665kcal/mol), and 0.713 (1.0kcal/mol) for hydrophobic, polar, negative-charged, positive-charged, alanine, and non-alanine respectively. Note that consistent results are achieved except the mutation type of positively-charged, which demonstrates our model's ability to generalize across diverse amino acid substitutions without strong bias. The relatively lower performance observed for positively-charged mutations may be attributed to the small number of samples in this group, which includes only 57 mutations. This limited data size may reduce the model’s ability to capture reliable patterns for this category.  

\paragraph{Performance Analysis based on Mutation Regions} 
We group the mutation sites into five structural regions: interior, surface, support, rim, and core. These regions are determined by their relative accessible solvent area (rASA) using specific cutoff values \cite{levy2010simple} as detailed in Table 1 of the supplementary information. Among these regions, mutations occurring in the core exhibit the highest average experimental $\Delta\Delta G$ value of 0.995 kcal/mol, suggesting a greater destabilizing effect upon mutation in this buried region. The comparison between experimental energy changes and predicted ones across these regions is shown in Fig. \ref{fig:protein-rna-result}\textbf{f}.
The corresponding PCC and MAE values are as follows: 0.83 (0.594 kcal/mol) for interior, 0.484 (0.677 kcal/mol) for surface, 0.815 (0.885 kcal/mol) for support, 0.613 (0.83 kcal/mol) for rim, and 0.727 (0.79 kcal/mol) for core regions.
The interior and support regions have the highest PCC, while the surface region has the lowest PCC. A possible explanation for the lower correlation in the surface region is the limited variability in its experimental data, which exhibits the smallest variance in $\Delta\Delta G$ values (1.05 kcal$^2$/mol$^2$). This reduced variance may hinder the model’s ability to capture meaningful trends, thereby decreasing correlation despite reasonable error magnitudes.

\subsubsection{Performance on Protein-DNA Interactions}
\begin{figure}[h]
    \centering
    \includegraphics[width=1\linewidth]{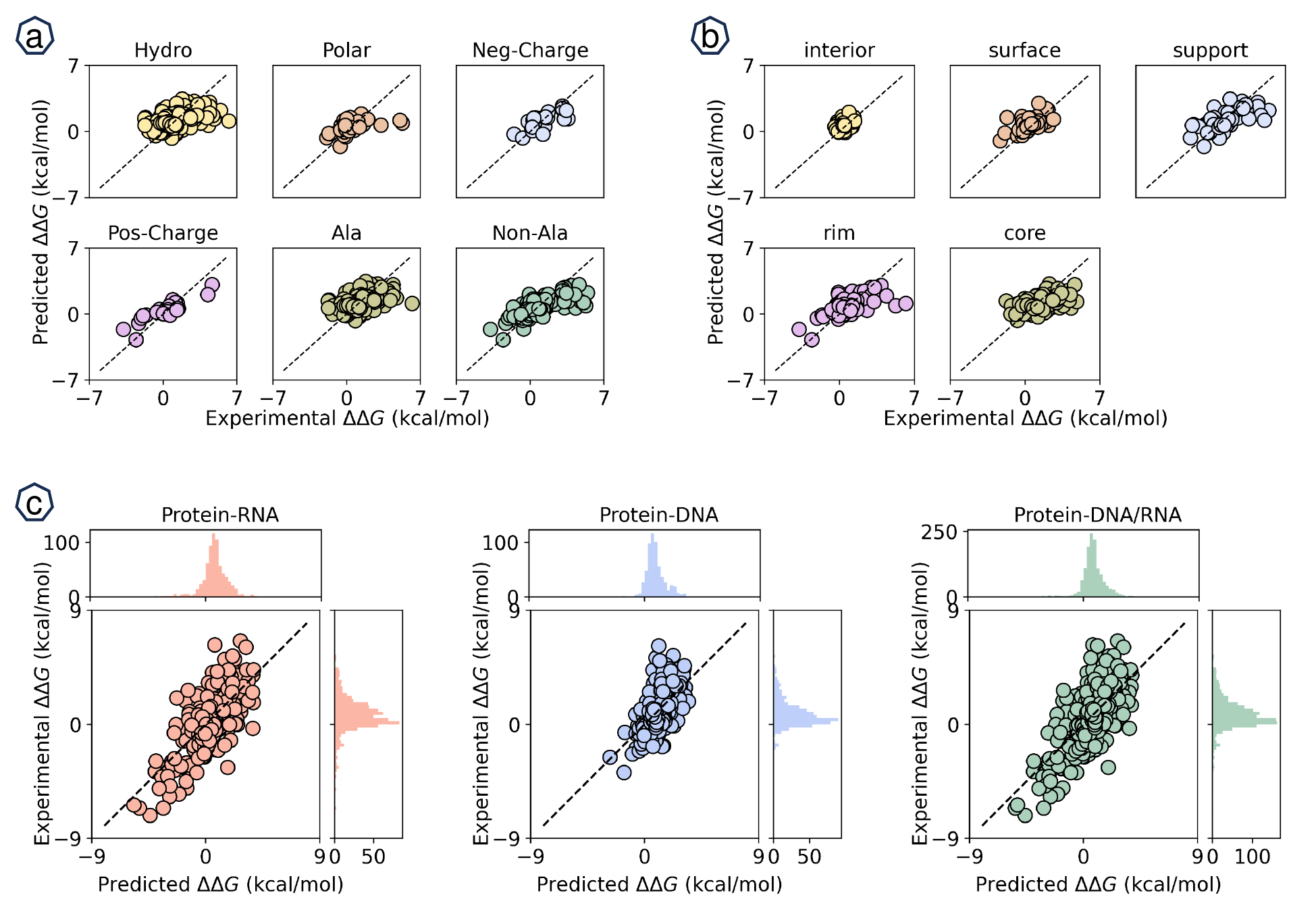}
    \caption{Illustration of model performance on predicting mutation-induced protein-DNA binding free energy changes for the S596 dataset. \textbf{(a)}:  Model performance across different mutation types based on residue physicochemical properties. \textbf{(b)}: Model performance across different structural regions defined by relative accessible solvent area (rASA). \textbf{(c)}: Comparison between experimental and predicted $\Delta\Delta G$ for protein-DNA interactions, protein-RNA interactions, and both of them. }
    \label{fig:protein-dna-result}
\end{figure}

For protein-DNA interactions, the S596 dataset is derived by combining the S419 and S117 datasets. Specifically, S419 was collected by the authors of the SAMPDI-3D model, while S117 is a newly curated dataset by the updated SAMPDI-3Dv2 model. The SAMPDI-3Dv2 model was trained on the S417 datset and evaluated on the S117 dataset, achieving a PCC of 0.17 and an RMSE of 1.34 kcal/mol. Following the same training and testing strategy, our model TopoML achieves a superior performance with a PCC of 0.332 and an RMSE of 1.23 kcal/mol. SAMPDI-3Dv2 also performed the 5-fold cross-validation on the S569 dataset, it obtained a PCC of 0.65 without an MAE reported. Under the same setting, our model achieved a PCC of 0.67 with an MAE of 0.61 kcal/mol. To obtain more robust and unbiased results, we further conducted 10-fold cross-validation, which yielded an improved PCC of 0.681 and an MAE of 0.612 kcal/mol.

Furthermore, We analyze the model performance across different mutation types based on the predictions from 10-fold cross-validation. The results are shown in Fig.  \ref{fig:protein-dna-result}\textbf{a}. Specifically, the PCC (MAE) values are 0.631 (0.597 kcal/mol) for hydrophobic, 0.498 (0.700 kcal/mol) for polar, 0.745 (0.644 kcal/mol) for negatively charged, 0.877 (0.648 kcal/mol) for positively charged, 0.643 (0.582 kcal/mol) for alanine, and 0.693 (0.673 kcal/mol) for non-alanine mutations. Additionally, we evaluated model performance across different mutation regions, as shown in Fig. \ref{fig:protein-dna-result}\textbf{b}. The PCC and MAE values are 0.334 (0.477 kcal/mol), 0.617 (0.481 kcal/mol), 0.684 (0.812 kcal/mol), 0.685 (0.592 kcal/mol), and 0.638 (0.634 kcal/mol) for interior, surface, support, rim, and core regions respectively. The performance is consistent across most regions, except for the interior, where lower predictive accuracy may be attributed to the smaller variance of experimental $\Delta\Delta G$ values (variance of only 0.241 kcal$^2$/mol$^2$). Note that for protein-DNA interactions, our model achieves a PCC of 0.681 and an MAE of 0.876 kcal/mol using 10-fold cross-validation. When combining the 596 protein-DNA and 710 protein-RNA mutations into a unified dataset of 1306 mutations, our model achieves a PCC of 0.712 and an MAE of 0.708 kcal/mol, further demonstrating the robustness of our model across interaction types. A comparison between the experimental and predicted $\Delta\Delta G$ values is shown in Fig. \ref{fig:protein-dna-result}\textbf{c}.

\subsubsection{Feature Analysis}
In our TopoML model, we incorporate three types of features: topological features derived from persistent Laplacians, sequence features obtained from a pretrained Transformer model, and physicochemical features. To evaluate the contribution of each feature type, we construct several ablation models using different combinations of these features. We denote the topological, sequence, and physicochemical features as Topo, Seq, and Phy, respectively. To ensure the reliability of the results, we perform 10-fold cross-validation on both the protein-DNA and protein-RNA interaction datasets. The performance of each feature combination is summarized in Table 
\ref{tab:feature-analysis}.

\begin{table}[h]
\centering
\caption{Performance comparison of different types of features. The unit for MAE is kcal/mol. Topo: topological feature from persistent Laplacian, Phy: physicochemical feature, Seq: sequence feature by pretrained Transformer. }
\label{tab:feature-analysis}
\begin{tabular}{|l| cc| cc| cc|cc| }
    \hline
     & \multicolumn{2}{c|}{Topo} & \multicolumn{2}{c|}{Phy} & \multicolumn{2}{c|}{Seq} & \multicolumn{2}{c|}{Topo+Phy+Seq} \\
    & PCC   & MAE   & PCC   & MAE   & PCC   & MAE   & PCC   & MAE   \\
    \hline
    Protein-DNA & 0.648 & 0.65 & 0.638 & 0.654 & 0.608 & 0.657 & 0.681 & 0.612\\
    \hline
    Protein-RNA & 0.678 & 0.825&  0.67 & 0.843 & 0.678 & 0.817 & 0.721 & 0.785\\
    \hline
\end{tabular}
\end{table}

For protein-DNA interactions, the topological features yield the best performance, achieving a PCC of 0.648 and an MAE of 0.65 kcal/mol. In comparison, models using only physicochemical and sequence features achieve PCCs of 0.638 (MAE: 0.654 kcal/mol) and 0.608 (MAE: 0.657 kcal/mol), respectively. For protein-RNA interactions, all three feature types result in similar performance levels.
These findings highlight the importance of topological features in capturing essential information, especially for protein-DNA interactions. Moreover, integrating all three feature types leads to a more robust and generalizable model performance.


\section{Conclusions}
Studying the effects of protein mutations on protein-nucleic acid binding affinities is essential for understanding disease mechanisms and developing effective therapeutic strategies. However, experimental methods for measuring binding affinities are often time-consuming and labor-intensive, while existing computational approaches yield only moderate predictive performance. Therefore, it is imperative to explore novel methodologies that can improve prediction accuracy. In this work, we propose a topological machine learning model that leverages the persistent Laplacian from topological data analysis to predict protein-nucleic acid binding affinity changes induced by amino acid mutations. By integrating features from multiple perspectives, including physicochemical properties, topological structures, and sequence embeddings derived from a pretrained protein Transformer, our model achieves a comprehensive and robust representation of protein-nucleic acid interactions. This integrative approach enables our model to outperform existing methods on both protein-DNA and protein-RNA datasets.

Our model can be further improved along the following lines:
(1) Topological features: In the current study, we employ the standard persistent combinatorial Laplacian of simplicial complexes. However, recent developments have introduced more advanced mathematical tools, such as the persistent sheaf Laplacian \cite{wei2025persistent2} and the persistent Dirac operator \cite{ameneyro2024quantum,suwayyid2024persistent}, which may provide deeper insights into the structure of protein-nucleic acid complexes. Exploring the applicability of these tools is a promising direction for future work.
(2) Machine learning algorithms: While we use the gradient boosting tree for regression in this study, alternative algorithms such as random forests \cite{breiman2001random} and XGBoost \cite{chen2016xgboost} may also offer competitive performance due to the small data sizes. Additionally, the stacking ensemble strategies that combine multiple learning algorithms typically yields performance better than any single one of the models \cite{wolpert1992stacked}, and thus merit further investigation.

\section{Materials and Methods}
Here, we describe three types of features extracted from protein-nucleic acid complexes. A summary of the software packages used in this study is provided in Section 3 of the Supplementary Information, and the detailed machine learning parameters are listed in Table 2 of the Supplementary Information.
\subsection{Topological Features}
\subsubsection{Persistent Laplacian of Simplicial Complexes}
\begin{figure}[h]
    \centering
    \includegraphics[width=1\linewidth]{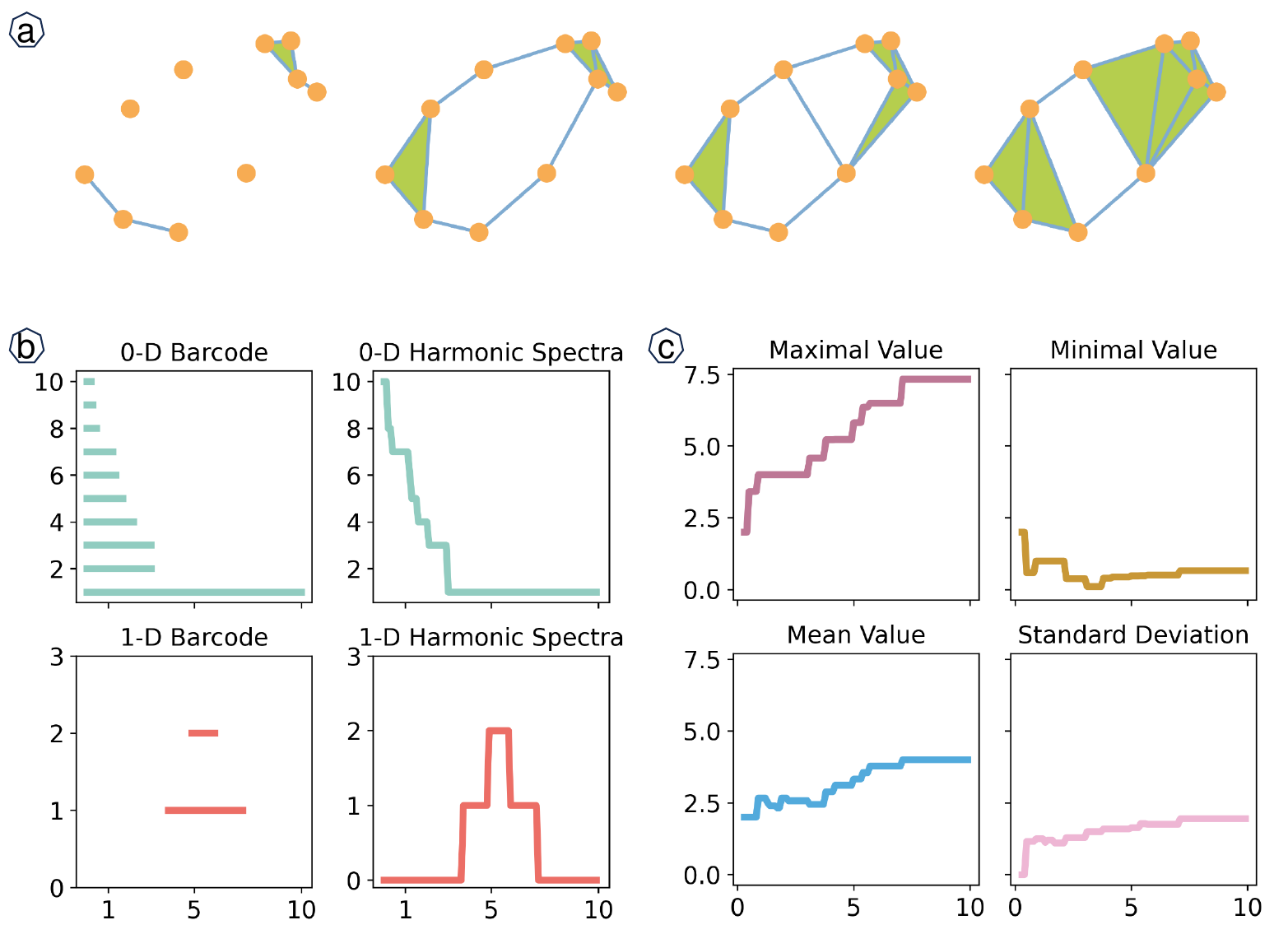}
    \caption{Illustration of persistent Laplacians. \textbf{(a)}: a filtration process of the Alpha complex from a point cloud data. \textbf{(b)}: the persistence barcodes of zero- and one-dimensional persistent homology (left), and the persistent multiplicity of zero eigenvalues from the Laplacian matrices in corresponding dimensions (right). \textbf{(c)}: persistent statistical attributes, including max, min, mean, and std, of the non-zero eigenvalues from the zero-dimensional Laplacian matrices.}
    \label{fig:persistent-laplacian}
\end{figure}
An abstract simplicial complex $K$ over a vertex set $V$ is a collection of non-empty subsets of $V$ such that for every $\sigma \in K$ and every nonempty subset $\tau \subset \sigma$, we have $\tau \in K$. An element $\sigma \in K$ consisting of $p+1$ vertices is called a $p$-simplex. The dimension of a $p$-simplex is defined as $p$, and the dimension of a simplicial complex is defined as the maximal dimension of its simplices. Any graph can be regarded as a one-dimensional simplicial complex. 

An oriented simplicial complex $K$ is a simplicial complex equipped with an ordering of its vertex set. Each simplex of $K$ inherits an orientation from the vertex ordering; a simplex together with this induced order is referred to as an oriented simplex. For simplicity, we refer to oriented simplices simply as simplices throughout the rest of this exposition.

We fix a field coefficient $\mathbb{F}$. For a simplicial complex $K$, the $p$-th chain group $C_p(K)$ is defined as the vector space over $\mathbb{F}$ with basis given by all the $p$-simplices in $K$. The boundary operator $\partial_p: C_p(K) \rightarrow C_{p-1}(K)$ is a linear map defined on a $p$-simplex $\sigma = [v_0 v_1 \cdots v_p]$ by
$$\partial(\sigma)=\sum_{i=0}^p(-1)^i[v_0v_1\cdots v_{i-1}v_{i+1}\cdots v_p].$$
These boundary operators satisfy the property $\partial_{p} \circ \partial_{p+1} = 0$, forming a chain complex:
$$\cdots\rightarrow C_{p+1}(K)\xrightarrow{\partial_{p+1}}C_{p}(K)\xrightarrow{\partial_p}C_{p-1}(K)\xrightarrow{\partial_{p-1}}\cdots\xrightarrow{\partial_1}C_0(K)\rightarrow0.$$
The $p$-th Betti number, denoted $\beta_p$, is the rank of $H_p(K)$. Intuitively, $\beta_0$ counts the number of connected components, $\beta_1$ the number of loops, and $\beta_2$ the number of voids. These topological invariants capture essential structural information about the simplicial complex.

Consider the standard inner product on the chain groups, that is, for any two simplices $\sigma$ and $\tau$,
$$<\sigma,\tau>=\delta_{\sigma,\tau},$$
with $\delta_{\sigma,\tau}$ being the Kronecker delta.
Let $\partial_p^*$ denote the adjoint of $\partial_p$ under this inner product. The $p$-th combinatorial Laplacian is then defined as
$$L_p=(\partial_p)^*\partial_p+\partial_{p+1}(\partial_{p+1})^*,$$
In matrix form, $\partial_p^*$ is the transpose of $\partial_p$. A fundamental property of the Laplacian is that the nullity of $L_p$ equals the $p$-th Betti number $\beta_p$.

Now consider a filtration of the simplicial complex $K$, i.e., a nested sequence of simplicial complexes:
$$K_0\subset K_1\subset \cdots \subset K_n=K.$$
Applying the homology functor to this sequence yields a sequence of homology groups:
$$H_p(K_0)\xrightarrow{i_0} H_p(K_1)\xrightarrow{i_1} \cdots \xrightarrow{i_{n-1}} H_p(K_n),$$
where $i_p$ is the homomorphism induced by the inclusion map from $K_p$ to $K_{p+1}$. Let $i_{s,t}$ denote the composition $H_p(K_s)\xrightarrow{i_s}H_p(K_{s+1})\xrightarrow{i_{s+1}}\cdots\xrightarrow{i_{t-1}}H_p(K_t)$. 

The $p$-th persistent homology from $K_s$ to $K_t$ is defined as
$$H_p^{s,t}=Im(i_{s,t}),$$
which are the homology classes that appear at $K_s$ and are still alive at $K_t$.

Let $C_p^s = C_p(K_s)$. Consider the subspace 
$$C_p^{s,t}=\{\alpha\in C_p^t|\partial(\alpha)\in C_{p-1}^s\}.$$
Then we have
$$
C_{p+1}^{s,t}\xrightarrow{\partial^{s,t}_{p+1}}C_p^s\xrightarrow{\partial_p}C_{p-1}^s
$$
where $\partial_{p+1}^{s,t}$ is the restriction of $\partial_{p+1}$ on the subspace $C_{p+1}^{s,t}$

The persistent Laplacian from $K_s$ to $K_t$ is defined as
$$L_p^{s,t}=\partial_{p+1}^{s,t}(\partial_{p+1}^{s,t})^*+(\partial_p)^*\partial_p.$$
Notably, the nullity of $L_p^{s,t}$ is the $p$-persistent Betti number from $K_s$ to $K_t$,
$$rank(H_p^{s,t})=rank(L_p^{s,t})$$
Consequently, the persistent homology information is fully encoded in the persistent Laplacian, which corresponds to its harmonic component. 

An example can be found in Fig.~\ref{fig:persistent-laplacian}. For the point cloud data consisting of 10 points, the resulting Alpha complex filtration process is shown in \textbf{(a)}. The corresponding persistent Laplacians are computed and presented in \textbf{(b)}: the persistence barcodes are shown on the left, and the persistent multiplicities of zero eigenvalues of the Laplacian matrices are shown on the right. The two bars in the 1-D barcode and the two zero eigenvalues in the 1-D harmonic spectrum correspond to the two loop structures in the simplicial complex representation. Notably, the barcode of persistent homology and the harmonic spectra of persistent Laplacian reflect the same curve information, illustrating the isomorphism theorem mentioned above. In addition to the harmonic spectra, the non-harmonic spectral information from the persistent Laplacians reveals additional geometric and combinatorial properties of the underlying data. The persistent statistical attributes, including the maximum, minimum, mean, and standard deviation of the non-zero eigenvalues, are shown in \textbf{(c)}.

\subsubsection{Persistent Laplacian for Protein-Nucleic Acid Interactions}
In our TopoML model, each protein-nucleic acid complex is represented as a set of simplicial complexes and their persistent Laplacians are computed to extract topological features. Since protein-nucleic acid complexes are typically of large sizes, and their interactions primarily occur at the interface regions, we restrict our analysis to atoms near the binding and mutation sites in order to reduce computational cost. More specifically, for each protein-nucleic complex, the following four atom sets are considered:
\begin{itemize}
    \item $\mathcal{A}_m$: atoms within the mutation site
    \item $\mathcal{A}_{mn}(r)$: atoms in the neighborhood of the mutation site within a cutoff of $r$
    \item $\mathcal{A}_p(r)$: protein atoms within a distance $r$ from the binding site
    \item $\mathcal{A}_n(r)$: nucleic acid atoms within a distance $r$ from the binding site
\end{itemize}
Both the wild-type and mutant protein structures are considered, and the element-specific representation \cite{cang2018integration} is used to capture pairwise interactions between different atom types. Specifically, the carbon (C), nitrogen (N), and oxygen (O) atoms in the mutation site and its neighborhood are used to form nine pairwise atom combinations. Similarly, the C, N, and O atoms in the protein and nucleic acid binding regions yield another nine combinations. In total, 36 atom sets are defined for each protein-nucleic acid complex. We construct both Vietoris-Rips complexes \cite{vietoris1927hoheren} and Alpha complexes \cite{edelsbrunner2011alpha} on these atom sets to model their topological interactions.

For Vietoris-Rips complex, the following distance $d$ is used to model the interactions
\begin{equation}
    d(x,y)=\begin{cases}
        E_d(x,y)&,S_x\ne S_y\\
        \infty&,otherwise\\
    \end{cases}
\end{equation}
where $E_d(x,y)$ denotes the Euclidean distance between atoms $x$ and $y$, and $S_x$ indicates the atom set to which $x$ belongs. For example, in modeling interactions between the mutation site and its neighboring atoms, any pair of atoms within the mutation site itself are considered to be at infinite distance. For each Rips complex, we compute the zero-dimensional persistent Laplacian. The harmonic component is calculated over a filtration from 0 \AA~ to 8 \AA~ with a step size of 0.5 \AA. A bin-counting method is applied to generate a 16-dimensional feature vector for each atom set. For the non-harmonic components, the same filtration is used, and seven statistical properties are extracted from the non-zero eigenvalue spectrum of the persistent Laplacian: maximum, minimum, mean, sum, standard deviation, variance, and count of eigenvalues. This yields a 112-dimensional feature vector per atom set.

For Alpha complex, the standard Euclidean distance is used for all 36 atom sets. In this case, we consider the one-dimensional and two-dimensional persistent Laplacians. To simplify computation, only the harmonic component is retained, which corresponds directly to persistent homology. Using barcode representations, we extract the following statistical features: (1) sum, maximum, and mean of bar lengths; (2) maximum and minimum of bar birth times; and (3) maximum and minimum of bar death times. This yields a 14-dimensional feature vector per atom set. In addition to the 36 atom sets, we also compute the same 14-dimensional Alpha complex features on two global atom sets: all heavy atoms in the wild-type and mutant complexes, respectively, to capture the global topological structural information. The total topological feature for a protein-nucleic complex is a 5140=(112+16)$\times$36+14$\times$38 dimensional vector. 
 
\subsection{Physicochemical Features}
In addition to topological features, our model incorporates a set of physicochemical features at both the atom-level and the residue-level to comprehensively characterize each protein-nucleic acid complex. At the atom-level, the features include solvent-excluded surface area, partial atomic charges, Coulombic interaction energies, van der Waals interaction energies, and electrostatic solvation free energy \cite{chen2011mibpb,rocchia2001extending,jurrus2018improvements}. At the residue-level, the features comprise amino acid composition in the neighborhood of the mutation site, pK$_a$ shifts due to mutation, Position-Specific Scoring Matrix (PSSM)-based features, and secondary structure-based descriptors. These features collectively form a 783-dimensional vector representing the physicochemical profile of each protein-nucleic acid complex. Details can be found in  Section 2 of the Supplementary Information.

\subsection{Sequence Features}
Protein large language models, such as ESM (Evolutionary Scale Modeling) \cite{rives2021biological} and ProtTrans \cite{elnaggar2021prottrans}, which are trained on hundreds of millions of protein sequences, have demonstrated remarkable performance in various protein analysis tasks. Their effectiveness has been further enhanced through the application of hybrid fine-tuning strategies. In the TopoML model, we employ the ESM-2 Transformer architecture \cite{lin2023evolutionary} to generate sequence embeddings. Specifically, we use the esm2\_t33\_650M\_UR50D model, which is trained using self-supervised learning via masked language modeling (MLM) on a comprehensive protein sequence database.
The ESM-2 model comprises 33 Transformer layers and approximately 650 million parameters. Each layer encodes input tokens into 1280-dimensional vectors. For each input sequence, we extract the representations from the final layer and compute the average across the sequence length, resulting in a single 1280-dimensional vector. Sequence embeddings are obtained for both the wild-type and the mutant protein sequences, and these vectors are concatenated to form a final 2560-dimensional representation used as the sequence feature in our model. For more details on the pretrained ESM-2 model, please refer to the reference \cite{lin2023evolutionary}.

\section*{Data and Code Availability}
The data and code in this study can be found in  \href{https://github.com/LiuXiangMath/TopoML}{github.com/LiuXiangMath/TopoML}.

\section*{Acknowledgment}
This work was supported in part by NIH grants R01GM126189, R01AI164266, and R35GM148196, National Science Foundation grants DMS2052983 and IIS-1900473, Michigan State University Research Foundation, and  Bristol-Myers Squibb 65109.

\section*{Conflict of Interest}
The authors declare no conflict of interest.



\newpage
\section*{ Supplementary Information  }


\setcounter{section}{0}
\section{Structural Region Types}

\begin{table}[h]
\centering
\caption{Structural regions definition based on rASA, the definition is from \cite{levy2010simple}. rASAm: relative ASA in monomer; rASAc: relative ASA in complex; $\Delta \rm rASA=rASAm-rASAc$.  }
\label{tab:region-type}
\begin{tabular}{|l| c| c| c| }
    \hline
    Region Type & $\Delta rASA$ & rASAc & rASAm  \\
    \hline
    Interior& 0 & $<$0.25 & \\
    \hline
    Surface& 0 & $>$0.25 & \\
    \hline
    Support & $>$0 & & $<$0.25\\
    \hline
    Rim & $>$0 & $>$0.25 &\\
    \hline
    Core & $>$0& $<$0.25&$>$0.25\\
    \hline
\end{tabular}
\end{table}

\section{PhysicoChemical Features }
In TopoML, in addition to topological and sequence features, we incorporate physicochemical features to capture crucial chemical and physical properties for protein-nucleic acid interactions. These physicochemical features are divided into two major categories: atom-level and residue-level features. The inclusion of these features ensures a comprehensive and robust characterization of the protein-nucleic acid interactions. The total feature dimension for the physicochemical component is 783.
\subsection{Atom-level Features}
For atom-level features, the element-specific strategy is employed to categorize atoms into distinct sets. Specifically, for proteins, atoms are grouped into seven categories: carbon (C), nitrogen (N), oxygen (O), sulfur (S), hydrogen (H), heavy atoms, and all atoms. Similarly, for nucleic acids, atoms are grouped into carbon (C), nitrogen (N), oxygen (O), phosphorus (P), hydrogen (H), heavy atoms, and all atoms. These groups are further subdivided based on spatial regions, including the mutation region, the mutation neighborhood region, the protein binding region, the nucleic acid binding region, and the entire structure. The atom-level features are generated based on these groups. The wild-type, mutant, and their difference are considered to give a comprehensive representation.
\begin{enumerate}
    \item Surface Area: Atom-level solvent-excluded surface areas are calculated using ESES \cite{liu2017eses}. Atom surface areas within each group are summed up to generate one feature, leading to $(5\times 7+1)\times 3=108$ features.
    \item Partial Charge: Partial charges are computed using PDB2PQR \cite{dolinsky2004pdb2pqr} with the AMBER force field. For each atom group, both the sum of partial charges and the sum of absolute partial charges are considered. This leads to $5\times 7\times 3 \times 2+2\times 3=216$ features.
    \item Coulomb Interaction: The Coulomb energy of the $i$-th atom is computed as the sum of pairwise Coulomb interactions with every other atom according to the following formula:
    $$C_i=\sum_{j\ne i}k_e\frac{q_iq_j}{r_{ij}}$$
    where $k_e$ is the Coulomb constant. The constant of 1 is used in our computation. $q_i$ is the partial charge of the $i$-th atom, and $r_{ij}$ is the Euclidean distance between $i$-th and $j$-th atoms. Only five atom element groups (C, N, O, S/P, and heavy atoms) are considered in the computation. Both the coulomb energy and its absolute value are computed. Totally, there are $4\times 5\times 2\times 3+6\times 2\times 3=156$ features.
    \item Van der Waals interaction. The van der Waals energy of the $i$-th atom is computed as the sum of pairwise Lennard-Jones potentials with all other atoms.
    $$V_i=\sum_{j\ne i}\epsilon [(\frac{r_i+r_j}{r_{ij}})^{12}-2(\frac{r_i+r_j}{r_{ij}})^6]$$
    where $r_i$ is the atom radius of $i$-th atom and $\epsilon$ is the depth of potential well. We use $\epsilon=1$ in the computation. Similarly, only 5 atom element groups (C, N, O, S/P, and heavy atoms) are considered. The feature size is $4\times 5\times 3+6\times 3=78$.
    \item Electrostatic Solvation Free Energy. The electrostatic solvation free energy for each atom is computed using the Poisson-Boltzmann model through the software MIBPM \cite{chen2011mibpb}. By summing up all solvation free energies within each group, $5\times 7\times 3+3=108$ features are generated.
    \end{enumerate}
 
\subsection{Residue-level Features}
\begin{enumerate}
    \item Mutation Neighborhood Amino Acid Composition. Neighboring residues within 12 Å of the mutation site are considered. These residues are classified into six categories: hydrophobic, polar, polar uncharged, positively charged, negatively charged, and special cases. Both the count and percentage for each category are used as environmental features. Additionally, the sum, mean, and variance of residue volumes, surface areas, molecular weights, and hydropathy scores are calculated, along with the sum of charges. There are a total of $6\times 2+3\times 4+1=25$ features.
    \item pKa shifts. The pKa values of residues are computed using PROPKA software \cite{bas2008very}. Specifically, the seven ionizable amino acids are considered: ASP, GLU, ARG, LYS, HIS, CYS, and TYR. The pKa values of the mutation site, N-terminal, and C-terminal residues are considered for both wild-type and mutant types. The maximum, minimum, and sum of the differences between the wild-type and mutant pKa values are computed, along with the minimum and sum of their absolute differences. Additionally, the sum and absolute sum of pKa values within each group are included. This results in $3\times2+3+2+7\times 2=25$ features.
    \item Position Specific Scoring Matrix. Features are computed from the conservation scores in the PSSM of the mutation site. The conservation scores are computed by PSI-BLAST \cite{altschul1997gapped}. The feature size is $8\times 2=16$ for wild-type and mutant proteins.
    \item Secondary Structure. The SPIDER software \cite{yang2017spider2} is used to compute the probability of residue torsion angle and a residue being in a coil, alpha helix, and beta strand. The features are computed for wild-typ, mutant type, and their difference, resulting in $17\times 3=51$ features.
\end{enumerate}

\section{Software Packages}
\begin{itemize}
    \item The Jackal software \cite{xiang2002jackal} is used to generate mutant protein-nucleic acid complex. 
    \item The PDB2PQR \cite{dolinsky2004pdb2pqr} with the AMBER force field is used to generate the partial charges of atoms.
    \item The ESES software \cite{liu2017eses} is used to compute the solvent excluded surface area of atoms. 
    \item The MIBPB \cite{chen2011mibpb} is used to compute the electrostatic solvation free energy of atoms.
    \item The PROPKA package \cite{bas2008very} is used to compute the pka values.
    \item The PSI-BLAST software \cite{altschul1997gapped} with an uniref50 database is used to compute the PSSM.
    \item The SPIDER software \cite{yang2017spider2} is used to compute the secondary structure features.
    \item The GUDHI \cite{maria2014gudhi} and scipy \cite{gommers2024scipy} packages are used to compute the persistent Laplacian features.
    \item The ESM-2 \cite{lin2023evolutionary} is used to generate the sequence features.
\end{itemize}

\section{Machine Learning Parameters}

\begin{table}[h]
\centering
\caption{Parameters for gradient boosting tree  }
\label{tab:region-type}
\begin{tabular}{|c| c| }
    \hline
    No. of estimators &  10000 \\
    \hline
    max\_depth & 7\\
    \hline
    min\_sample\_split & 3 \\
    \hline
    learning\_rate & 1e-3\\
    \hline
    max\_features & sqrt\\
    \hline
    subsample & 0.5\\
    \hline
\end{tabular}
\end{table}

\end{document}